\documentclass[runningheads]{llncs}
\usepackage[T1]{fontenc}
\usepackage{geometry}
\usepackage{hyperref}
\usepackage{graphicx}
\usepackage{xcolor}
\usepackage{amsmath}
\usepackage{cleveref}
\usepackage{booktabs}
\usepackage[breakable]{tcolorbox}
\usepackage{algorithm}
\usepackage{algorithmicx}
\usepackage{algpseudocode}
\usepackage{subcaption}

\begin{document}
\title{From Caricature to Collective Artifact: How Persona Construction Drives Polarization in Generative Social Simulations}

%
\titlerunning{From Caricature to Collective Artifact}
%
\author{Dimitris Michailidis\inst{1}\orcidID{0000-0002-0106-1126} \and Petter Törnberg\inst{2} \orcidID{0000-0001-8722-8646} \and
Fernando P. Santos \inst{1}\orcidID{0000-0002-2310-6444}}
\authorrunning{D.Michailidis, P. Törnberg, F.P. Santos}
%
\institute{Socially Intelligent Artificial 
Systems, University of Amsterdam \and Institute for Logic, Language and Computation, University of Amsterdam \\
\email{\{d.michailidis, p.tornberg, f.p.santos\}@uva.nl}}

\maketitle              
\begin{abstract}
Generative Social Simulation (GSS) leverages Large Language Models (LLMs) to create sophisticated personas that define agents with demographics, emotions, and political opinions. While prior work shows that personas compress outputs into stereotypical caricatures, there is no comprehensive study of how these caricatures evolve into collective artifacts when they interact. Using a Twitter-like GSS platform with personas derived from the American National Elections Study (ANES), we ablate three layers of partisan identity: affective polarization markers, party identification, and vote choice, while retaining demographics, ideological positioning and important issues. We find that collective outcomes depend heavily on how survey responses are translated into personas. Removing a few explicitly partisan statements moves the network from near-complete segregation to cross-party ties approaching those of real Twitter networks. We argue that matching one outcome is not evidence for behavioral validity and point to salience as a possible explanation. Surveys ask for a position on every question regardless of its importance to the respondent, and persona construction turns each answer into a direct behavioral instruction. We conclude with a checklist for more comprehensive evaluation of personas in future works.

\keywords{Generative Social Simulation  \and Personas \and Mechanism Ablation \and Polarization.}
\end{abstract}
\section{Introduction}
Generative social simulation (GSS) uses large language models to simulate social behavior among artificial agents. A common application simulates synthetic social media platforms with agents that consume news, share content and form connections
\cite{gao2023s3,tornberg2023simulatingsocialmediausing,rossetti2024social,larooij2025can,orlando2025can,zhang2025ga}. In these simulations, a common approach is to define agents through a textual \textit{persona}: a description, often derived from survey data, that specifies the agent's demographics, beliefs, political leanings, and interests \cite{argyle2023out,ferraro2024agent,chang2025llms,larooij2025can,orlando2026validating}. Personas take the place of parameters in classical agent-based models, which have long been criticized for their simplicity. Instead of a homophily parameter, a GSS agent receives an instruction, say, "You strongly prefer to connect with like-minded others."

Personas are meant to introduce more nuanced and culturally relevant instructions. However, because agent behavior is influenced by the patterns in the text Large Language Models (LLMs) are trained on \cite{nudo2026generative},  personas can inherit biases present in the data. There is plenty of evidence that personas poorly represent the people they portray \cite{li2026llm}. They produce more stereotypical and marked language than real people \cite{cheng2023marked}, caricature the groups they are asked to portray \cite{cheng2023compost,nudo2026generative}, are biased toward American cultural norms \cite{cao2023assessing}, give very homogeneous output \cite{anthis2025position,wu2025llm,lukauskas2026plausible}, lack the emotional diversity of humans \cite{pagan2025computational}, and are more partisan than the humans they represent \cite{bisbee2024synthetic,jeoung2025examining}. These works compare individual persona answers against human ones \cite{lukauskas2026plausible}. How collective-level outcomes emerge when personas interact remains less understood.

Recent studies report that simulated social networks trend toward polarization, segregation, and hierarchy \cite{larooij2025can,orlando2025can,piao2025emergence}. Do these results reflect real social media platforms? Real-world evidence shows that pro-social behavior constantly emerges in online social networks \cite{dorr2025research}. If humans find common ground, why do their digital equivalents fail?

We study this question using political polarization as a case study. We argue that collective outcomes in GSS can be artifacts of how attributes are translated into personas, and show that even small changes in personas can alter those outcomes. Surveys ask respondents to take a position on every topic on a questionnaire, whether or not it matters to them, and record where they stand, but not necessarily how much they care \cite{vyalyh2021and}. Converting these responses into personas creates a flattened representation of a person \cite{wang2025large}: each answer becomes an always-active description, equally weighted with every other aspect of their personality. One possible explanation of our results is \textit{salience}: given the tendency of LLMs to output text consistent with the input \cite{nudo2026generative}, every political statement in the persona acts as an instruction. Even statements such as "You never talk about politics" can create an "ironic rebound" effect and activate the topic \cite{mann2025don}. We investigate how these individual-level distortions affect collective behaviors in artificial intelligence systems \cite{rahwan2019machine}, particularly the emergence of polarization.

We ask three research questions. \textbf{RQ1}: do personas amplify the political attitudes of the respondents they represent, even without any social interaction between them? We use a standard affective-polarization questionnaire under increasingly ablated persona conditions and compare their responses with the original American National Election Studies survey (ANES) responses. \textbf{RQ2}: do persona-level distortions translate into collective network artifacts? We perform controlled simulation ablations, removing partisan information from both public biographies and internal personas while holding the platform fixed. \textbf{RQ3}: does enriching personas with non-political statements reduce the effect of their partisan identity? We introduce non-political preferences and measure whether political polarization changes. Together, we provide a 
comprehensive individual and collective-level ablation of political personas in generative social simulations, and show how persona-level distortions become collective-level artifacts in the simulation.

Our experiments show that LLM agents are very sensitive to their assigned identities, and that this sensitivity
becomes a collective artifact at the network level. Full personas lead to extremely polarized networks, while ablated agents form cross-party ties more often, approaching those observed in empirical studies \cite{halberstam2016homophily,gopal2025national} and engage with less polarizing content, even without being stripped of all political orientation. Questionnaire answers show that agents are polarized before any interaction, and our simulations show that persona specification is sufficient to produce large differences in collective outcomes within the same platform. In simulations, ablating public biographies has little impact on collective outcomes, while ablating internal personas has significant impact, suggesting that internal personas are more consequential than how agents perceive each other. No ablation condition matches all empirical benchmarks, meaning that a single outcome is not enough evidence for behavioral validity. Finally, we show that enriching personas with non-political statements does not reduce polarization. We conclude with a concrete checklist for persona construction, ablation, and reporting in GSS.

\section{Related Work}
For comprehensive surveys of LLM-based agents in social simulation, we refer the reader to recent reviews \cite{wang2024survey,mou2024individual,bernard2026large}. Here, we focus specifically on simulated social media interactions.

\subsection{Generative Social Media Simulation}
Park et al. introduced the concept of LLM agents with memory, planning, and social reasoning, showing that they produce believable behavior in a sandbox environment \cite{park2023generative}. Ghaffarzadegan et al. formalized generative agent-based modeling (GABM) as a methodological framework \cite{ghaffarzadegan2024generative}.

Several systems have since been proposed to model platform-level phenomena. Gao et al.'s $S^3$ models users as LLM agents that read messages, update their emotional state, and reshare or create content, reproducing information, emotion, and attitude propagation \cite{gao2023s3}. Zhang et al. extend it with group agents \cite{zhang2025ga}. Yang et al. built OASIS, a large-scale Twitter and Reddit simulator capable of scaling to millions of agents, demonstrating information propagation and opinion herding dynamics \cite{yang2024oasis}. Other work combines LLMs with opinion dynamics models to study echo chamber formation \cite{gu2025large}, or simulates Twitter-like platforms with personas derived from real posts, replicating empirical phenomena such as the friendship paradox \cite{ferraro2024agent,mou2024unveiling,orlando2026validating}.
 
A closely related line of work examines generative social simulation (GSS) specifically, in which LLM agents operate in synthetic social media platforms and interact through posting, re-posting, and following. Tornberg et al. proposed one of the earliest such frameworks, simulating a Twitter-like environment where persona-driven agents consume and share news \cite{tornberg2023simulatingsocialmediausing}. Subsequent work has extended this approach by incorporating different algorithmic feed designs \cite{larooij2025can,piao2025emergence} and by studying whether the resulting networks trend toward polarization, segregation, and hierarchical structures \cite{orlando2025can}. Our experiments build directly on this GSS framework \cite{larooij2025can}.
 
Beyond platform-level simulations, research has shown that multi-agent LLM systems can reproduce properties of human social networks, such as preferential attachment, triadic closure, and small-world effects \cite{papachristou2025network}. LLMs also show stronger homophily than observed in the real world \cite{chang2025llms}.
 
While these systems vary in memory, reasoning, or representation, they share a common reliance on textual personas to drive agent behavior. We instead examine how sensitive simulation outcomes are to persona specifications.

\subsection{Evaluating Personas for social simulation}
Individual-level biases of persona-enhanced LLM agents are well documented (see Introduction). We ask how persona design affects collective outcomes in artificial systems, an endeavor closely aligned with understanding dynamics of machine behaviour \cite{rahwan2019machine}. Recent works examine this at the individual level. Gu et al. find that LLM-generated social media posts have less linguistic diversity than real-world posts \cite{gu2025large}. Cheng et al. introduced CoMPosT, one of the first frameworks to identify and evaluate the caricature problem in LLM simulations with personas \cite{cheng2023compost}. We extend this work by demonstrating empirically what happens when such caricature personas are embedded in dynamic social network simulations.
 
At the network level, Chang et al. prompt LLMs to generate social networks in a single pass based on personas sampled from the US census, finding that LLMs substantially overestimate political homophily relative to empirical levels \cite{chang2025llms}. Their ablation study further reveals that political affiliation is the strongest predictor of homophily in generated networks, outweighing gender, age, race, and religion. Our setting differs in that we do not examine one-shot network generation, but embed personas in dynamic social simulations and show that internal persona instructions, and not just demographic labels, are the primary driver of polarized network outcomes.
 
Bisbee et al. also use the ANES survey to generate personas and prompt them to answer political feeling thermometer questions \cite{bisbee2024synthetic}. They find that, although average responses resemble human ones, subgroup-level analysis reveals higher partisanship in synthetic responses than in real data. We perform a more fine-grained ablation of the political components of the persona to identify the specific mechanisms that produce hyper-partisan synthetic users.
 
At the methodological level, Li and Tao argue that a fundamental mismatch exists between what agentic simulations optimize for and the demands of social simulation, particularly behavioral validity that is robust to implementation details \cite{li2026position}. They warn that personas are vulnerable to stereotyping and prompt sensitivity and call for mechanism ablations to test whether outcomes reflect emergent dynamics or artifacts of design choices. These concerns are systematized by Touzel et al., who identify common evaluation fallacies in LLM social simulations and argue for shared benchmarks and reproducible evaluation protocols \cite{touzel2026}. Ye et al. even demonstrate that small changes in writing style cascade into massive differences in macro-level interactions \cite{ye2026stop}. Salience has recently been investigated in generative agents, with researchers pointing to additional political context in personas as increasing ideological extremism \cite{nudo2026generative}. Their context, however, comes from users' previous posts on X rather than from surveys that are more representative of the general population. Our method also pinpoints which information increases polarization.

\section{Personas in Generative Social Simulation}

Personas are the core of agents in GSS. A scalar opinion parameter in the range $[-1, 1]$ in traditional ABMs translates to a qualitative statement such as "You strongly believe in X." Scalar values, however, carry both magnitude and direction. A modeler defines how much weight will scalar parameters have on the agent's update rules. In LLM-based agents, the model infers the relative weight of each statement in the prompt, which may give partisan statements more influence.

\begin{figure}[t]
    \centering
    \includegraphics[width=0.8\linewidth]{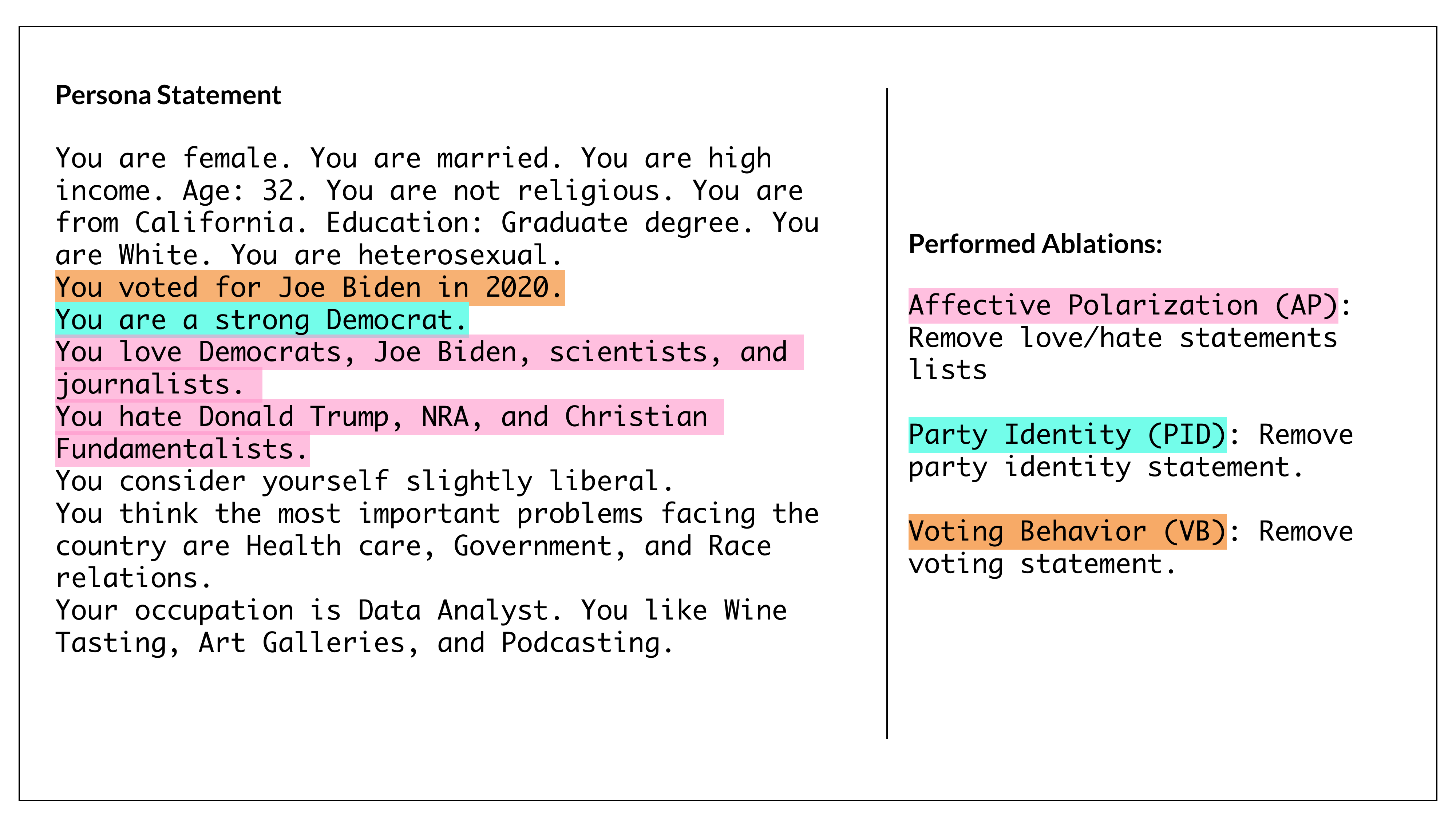}
    \caption{To evaluate how sensitive simulation outcomes are to persona specifications, we cumulatively ablate three layers of political identity: (1) affective polarization (AP), (2) party identification (PID) and (3) past voting behavior (VB) and simulate answering a polarization questionnaire and interacting in a social media platform. Complete persona statements in Appendix 1.2)}
    \label{fig:persona_ablations}
\end{figure}

Personas are built from survey data, like ANES \cite{anes_2020}, a four-yearly survey of demographics, political participation, party affiliation, voting behavior, and policy attitudes. By converting answers into statements, an agent takes up a very specific persona. \Cref{fig:persona_ablations} shows an example. Surveys do not measure importance to the respondent. Once converted to persona statements, all responsed become always active and equally important. This motivates our ablation study: rather than changing how much information a persona contains \cite{li2026llm}, we change the attributes that are stated, while holding the underlying infrastructure fixed.

\section{Methodology}
\label{sec:Methodology}

\subsection{Persona Construction}
We generate personas by randomly sampling respondents from the 2020 ANES \cite{anes_2020} (Details in Appendix 1.2). We extract demographics (gender, age, marital status, income, education, race, sexual orientation, religion, state), political engagement, ideology, most important problems, thermometer answers for political parties, organizations and movements (details in Appendix Table 3), party registration, voting behavior, interests, TV programs. Love/hate list statements are appended on the condition: if the thermometer answer is $\leq 10$, then it goes on the hate list, and if the thermometer answer is $\geq 90$, then it goes on the love list, in accordance with previous works \cite{tornberg2023simulatingsocialmediausing,larooij2025can}. We also generate a public social media biography by prompting the LLM-model with the full persona. Each agent, therefore, has two texts: the internal persona, which is included in the system prompt of every LLM call, is visible only to that agent and determines which posts to repost, headlines to share, and profiles to follow, and the public biography (max 140 characters), which is what other agents see about the author of a post when reposting it, before deciding whether to follow them (details and prompts in Appendix 1.4).

\subsection{Persona Ablation Design}
We ablate three layers of political identity in the agent persona:

\begin{enumerate}
    \item \textbf{Affective Polarization (AP)}: Removing love/hate statements about institutions or individuals (e.g., "You love X").
    \item \textbf{Party Identification (PID)}: Removing the statement "You are a strong/moderate Democrat/Republican/ Independent."
    \item \textbf{Voting behavior (VB)}: Removing the statement "You voted for X in 2020." Details on the statements in Appendix 1.2.
\end{enumerate}

These ablations are applied cumulatively: condition 1 removes AP, condition 2 removes AP \& PID, and condition 3 removes all three layers (AP, PID \& VB). The full-persona baseline retains all statements (\Cref{fig:persona_ablations}).

Even under condition 3, agents are not stripped of all political orientation, which would trivially reduce observed polarization. They retain their ideological identity (e.g., "You consider yourself slightly liberal"), their beliefs about important societal issues (e.g., healthcare, race relations), their demographics, and their personal interests.

We apply the ablations in two phases. First, we remove the statements only from the agent's public-facing biography, and then we remove them from both biography and internal persona, directly modifying the prompt that dictates each agent's behavior. More details in Appendix 1.3 and Appendix Table 5.

\subsection{Simulation Platform}
We use the GSS framework of \cite{larooij2025can}, to create a Twitter-like platform. At each timestep, an agent reads a feed of ten posts, through two strategies: \textit{chronological}: ten most recent posts, and \textit{boost out partisan}: five from other agents they follow and five selected by an algorithm, which prioritizes content from agents of the opposing political party \cite{larooij2025can}. The agent then chooses to repost one of the ten visible posts, or share a news headline from a curated database with an original comment, or take no action. If they repost, they are provided with the author's public biography and follower count and decide whether to follow them. These iterative interactions produce an evolving network that we can evaluate across several metrics. We provide the full pseudocode of the simulation and the prompts in Appendix Algorithm 1, and more details on the parameters in Appendix Table 1.

The feed algorithm, action space, news database, network initialization, and scheduling remain fixed. Only the textual content of the agent persona changes. This allows us to attribute differences in collective outcomes to persona specification rather than to the simulation infrastructure.

\subsection{Evaluation}
Before the simulation, each persona answers a standard three-part polarization questionnaire \cite{iyengar2019origins}. First, a feeling thermometer toward their own and the opposite party \cite{anes_2020}. Second, a series of specific trait questions, asking whether their own and the opposite party members are intelligent, honest, generous, hypocritical, selfish or mean \cite{levendusky2018americans,garrett2014implications}, and third, behavioral questions that assess whether each agent would follow others of the same/opposing party, or someone who loves or hates their or the opposing party's leader. Answers are stored and analyzed together with the original ANES responses. Full details on the prompts in Appendix 1.5.

Each question is asked in a separate API call, without shared history, with the full persona re-entered every time. Trait questions also allow a "don't know" response, so that the agents do not resort to a binary answer every time. Agents reply via an output schema, so responses are standardized. Following the static evaluation approach \cite{li2026position}, we assess the persona behavior through constrained tasks with measurable targets.

At the collective level, the \textit{mechanism ablation} addresses an important current credibility gap in generative social simulation: the lack of proper analysis on whether targeted edits lead to robust outcomes \cite{li2026position,touzel2026,ye2026stop}.

\subsection{Comparison with Empirical Networks}
We compare simulation outcomes with two observational studies of real political networks on Twitter \cite{halberstam2016homophily,gopal2025national}. Such datasets are rare because users' political orientation is hard to infer from their profiles. Halberstam and Knight study a network of 2.2 million political tweets during the 2012 US elections, classifying users as liberal or conservative based on how many politicians of each party they follow. They observe high homophily: only $33\%$ of liberal users follow conservatives, while $20\%$ of conservatives follow liberal users \cite{halberstam2016homophily}. Gopal et al. study a population of $4000$ US legislators on Twitter and find that $21.5\%$ of connections in total are cross-party (reporting only the overall rate). 

We note a few caveats about this comparison. First, neither network represents ordinary users: one consists of users who follow politicians and are therefore more likely to be strong partisans \cite{halberstam2016homophily}, the other of elected officials, whose incentives differ from everyday users \cite{gopal2025national}. Second, both datasets were collected during specific political periods (the 2012 elections and the COVID-19 pandemic in 2020, respectively), during which the partisan environments were different from the current or future ones. We use them as external benchmarks, not as a validation or a means to identify a correct persona configuration.

\subsection{Experimental Setup}


We simulate $500$ agents for $5000$ steps, with $10$ seeds per setting \footnote{The code for all experiments is available here: \url{https://github.com/dimichai/prosocialinterventions}}. We use OpenAI's \textit{gpt-5.4-mini}, \textit{gpt-4o-mini}, Google Gemini \textit{2.5-flash-lite} and Mistral's \textit{mistral-small-3.2-24b-instruct} for the LLM-driven components of our experiments. Due to its excessive cost, for \textit{gpt-5.4-mini} we sample $200$ personas and run the platform for $2000$ steps. Our focus is not to evaluate the technical performance of the models, but the design choices involved in employing them, and evaluate the robustness across models. Persona-induced biases have been shown to persist even when a model's analytical reasoning capabilities improve \cite{dash2025persona}, while model architecture and size have not been shown to affect results on textual consistency and fidelity \cite{pagan2025computational,samuel2025personagym}. More details on the models in Appendix Table 6.

\subsection{Metrics}
For the questionnaire, we report three metrics, each computed per agent and then averaged within party. To measure affective polarization, we use the thermometer gap, a standard metric that distinguishes warm feelings of the respondent's own party against the opposite party \cite{druckman2026measuring}.
\begin{equation}
    \Delta_{\mathrm{therm}} = T_{\mathrm{in}} - T_{\mathrm{out}},
\end{equation}
where $T_{in}$ is the mean $0$--$100$ rating of agents feelings toward their own party, and $T_{out}$  toward the opposite party.

To measure trait attitudes we use the trait gap: the share of positive/negative traits attributed to the same party minus the share attributed to the opposite party, ignoring "don't know" responses.
\begin{equation}
    \Delta_{K} = \bar{y}^{K}_{\mathrm{in}} - \bar{y}^{K}_{\mathrm{out}},
\end{equation}
where $\bar{y}^{K}_{t}$ is the share of traits in set $K$ that party-$p$ agents attribute to target $t$, computed separately for $K^{+} = \{\text{intelligent}, \text{honest},
\text{generous}\}$ and $K^{-} = \{\text{hypocritical}, \text{selfish}, \text{mean}\}$. A polarized population gives $\Delta_{K^{+}} > 0$ and $\Delta_{K^{-}} < 0$. 

Finally, we measure the follow gap, as the difference in probabilities of following in-party agents minus the probability of following out-party agents. Similarly, we compute probabilities of following those who love the respective party's candidate.
\begin{equation}
    \Delta_{\mathrm{follow}} = f_{\mathrm{in}} - f_{\mathrm{out}},
\end{equation}
\begin{equation}
    \Delta_{j} = f_{\text{loves } j} - f_{\text{hates } j}.
\end{equation}
Positive values of $\Delta_{\mathrm{follow}}$  indicate a preference for co-partisans. A polarized population gives $\Delta_{\text{Trump}} > 0$ among Republicans and $\Delta_{\text{Trump}} < 0$ among Democrats.

For the simulation, we use three metrics: E-I index, the average clustering coefficient, and the correlation between retweets and partisanship.

The E-I index measures the extent of homophily within the network by comparing the number of external links ($EL$) between agents of different parties (Democrats and Republicans) to the number of internal links ($IL$) within the same party. The index is defined as:
\begin{equation}
    EI = \frac{EL - IL}{EL + IL}
\end{equation}

Where $EI \in [-1, 1]$. A value of $-1$ indicates perfect homophily and complete segregation.

We also used the clustering coefficient to measure the networks' general tendency to form tightly knit clusters, regardless of party identity. For a node $v$ with degree $k_v$, the local clustering coefficient $C_v$ is:

\begin{equation}
    C_v = \frac{|\{e_{jk} : v_j, v_k \in N(v), e_{jk} \in E\}|}{k_v(k_v - 1)},
\end{equation}
where $N(v)$ is the set of neighbors of $v$, and $E$ are the edges.

We then calculate the network-wide average clustering coefficient (average of all nodes).

Finally, to quantify the degree to which partisanship translates to social capital, we calculate the correlation between an agent’s absolute partisan score and their total received retweets. Partisanship is calculated using the ANES survey "feeling thermometer" responses \cite{larooij2025can}. The partisan score $P$ of an agent is calculated as:

\begin{equation}
    P = \frac{\text{feelingRepublican} - \text{feelingDemocratic}}{100}
\end{equation}

where $P \in [-1, 1]$, with $-1$ indicating a strong Democrat and $+1$ a strong Republican. We use $|P|$, so that we measure partisan strength of either party.
\begin{figure}[t]
    \centering
    \includegraphics[width=0.8\linewidth]{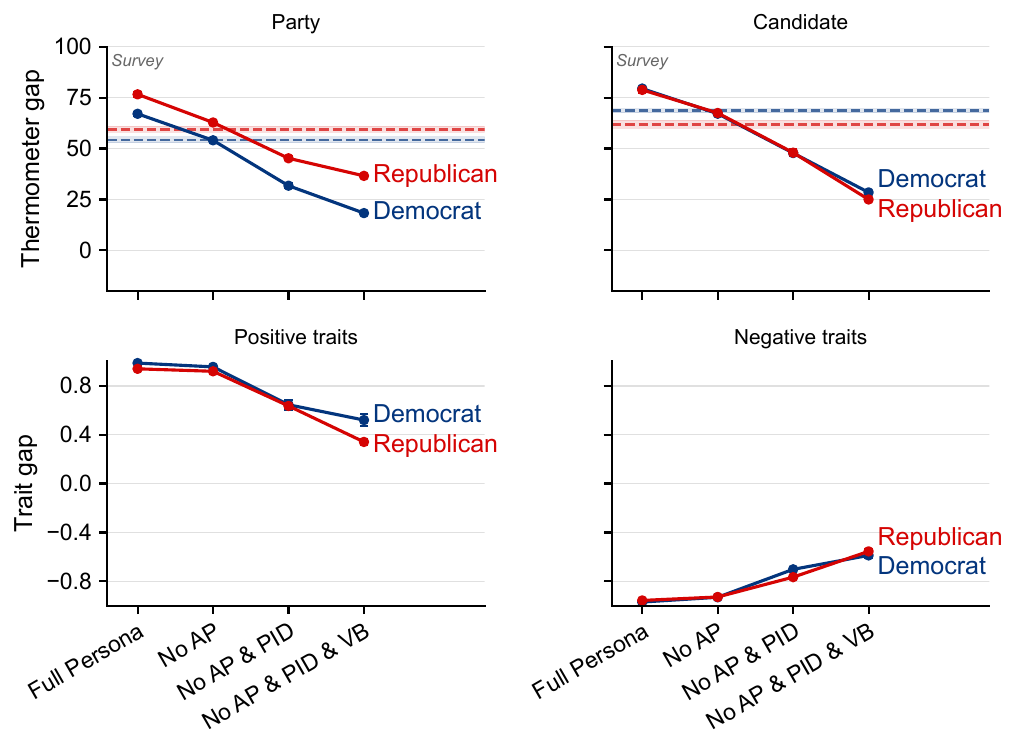}
    \caption{Thermometer (top) and trait (bottom) gaps by ablation condition. Dashed lines show the empirical gap from the ANES responses the personas were constructed from, with 95\% intervals. Personas amplify stated affective polarization. \textit{Top}: Full personas overestimate the survey by 15 points, while removing AP brings it to the ANES value. \textit{Bottom}: Under the full persona, agents attribute positive traits to their own party and negative traits to the other almost uniformly. Ablating AP has almost no effect; the gaps shrink only once PID is removed.}
    \label{fig:thermometer_trait_gap}
\end{figure}

\section{Results}
The results below use OpenAI's \textit{gpt-4o-mini} model and the \textit{boost out partisan} timeline algorithm. Our results remain consistent across LLM models and timeline strategies --- we fully report them in the Appendix Fig. 3. 

\subsection{RQ1: Persona ablations alter individual political responses before interaction}

Our goal is to observe if partisan markers are redundant for polarization, or if their removal eliminates the political information needed to reproduce it.

\textbf{Full personas amplify political polarization, while ablated ones dampen it}
\Cref{fig:thermometer_trait_gap} (top) shows the thermometer gap under each ablation, together with the original survey values when available. Under the full persona, agents show exaggerated polarization when asked both about how they feel about the parties and the candidates. Removing the affective polarization (AP) marker alone suffices to bring the agents to the ANES value. Further ablations move the agents below the observed benchmark, although not fully into neutrality.

\textbf{Trait gaps are reduced under further ablations}
\Cref{fig:thermometer_trait_gap} (bottom) shows the gaps in positive and negative trait attribution between in- and out-group agents (ANES has no comparable questions). Under the full persona, agents attribute every positive trait to their own party and every negative trait to the opposing one. All Democratic agents describe Republicans as hypocritical, selfish and mean, and vice versa. While AP affected the thermometer gaps, it has no effect on  personality traits. The gaps shift only once party identification (PID) is removed. Each persona layer affects the measure that corresponds to it: statements derived from thermometer items affect thermometer rating, while group-membership statements affect the agent's judgment of in- and out-group members.

\begin{figure}[t]
    \centering
    \includegraphics[width=1\linewidth]{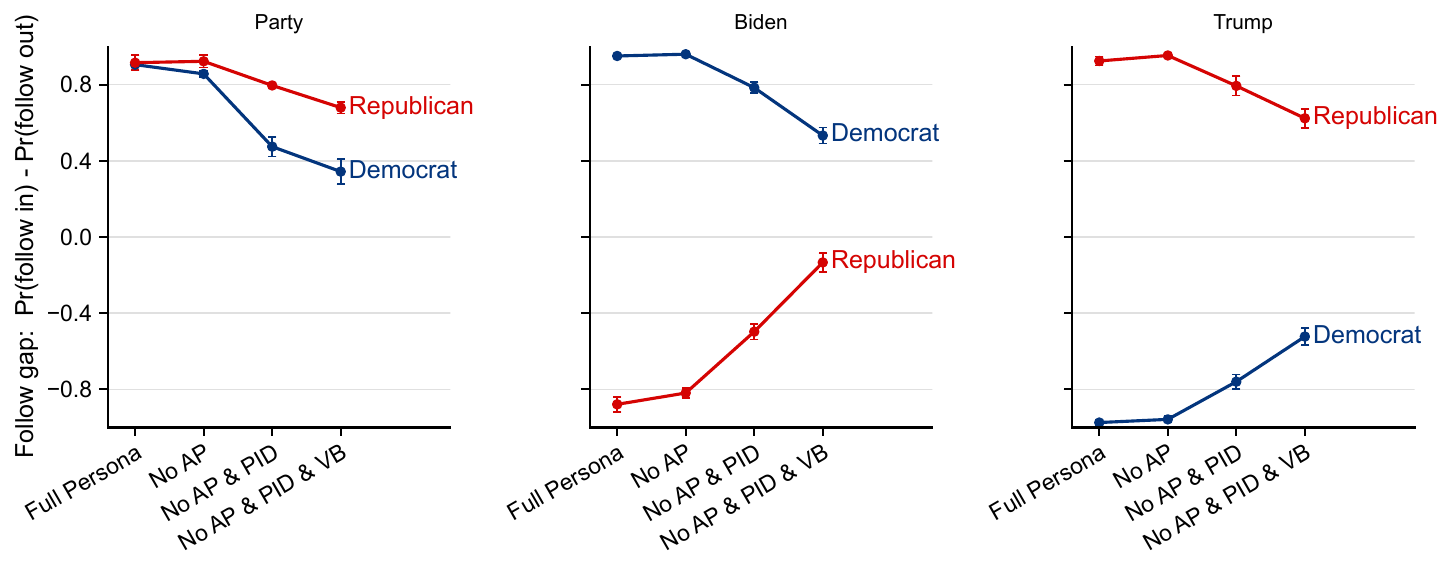}
    \caption{Follow gaps by ablation condition: party preference
(left) and preference between users who love and hate each candidate
(middle, right). Party identification dictates who agents say they will follow. Under the full persona, co-partisan preference is maximized and symmetric. Ablating AP leaves all three gaps unchanged. Further ablations reduce them and introduce an asymmetry: Republican agents are more reluctant than Democratic agents to follow the opposing party}
    \label{fig:follow_gap}
\end{figure}

\textbf{Party identification drives follow gaps}
\Cref{fig:follow_gap} shows that under the full persona the co-partisan preference is maximized and symmetric across parties. Here, too, removing the AP marker does not affect the gaps --- the decline begins only when PID is removed. The same pattern appears in the trait results, although the love/hate list might be expected to matter more in decisions about other groups. Additionally, we see an asymmetry introduced during the ablations. After removing PID and voting behavior (VB), Republican agents remain more reluctant to follow opponents than Democratic ones, with the gap between the two parties widening as more persona statements are removed. 

Under full ablation, $46.0\%$ of Democratic and $22.4\%$ of Republican agents report that they would follow a member of the opposing party, against the $33\%$ and $20\%$ empirically observed by Halberstam and Knight \cite{halberstam2016homophily}, while under the full persona both rates are much lower. While some observational studies have found some evidence of this asymmetry, \cite{halberstam2016homophily,boutyline2017social}, recent experimental work has shown that co-partisan preferences in link formation are symmetric across parties \cite{mosleh2021shared}. This finding is consistent with the broader finding that LLMs overestimate political homophily \cite{ferraro2024agent,chang2025llms} and exaggerate affective polarization \cite{bisbee2024synthetic}, and with the recent evidence that right-leaning users are more likely to be caricatured by LLMs than left-leaning ones \cite{nudo2026generative}.
  
Further ablations reduce extreme partisan behavior, but also reduce the gap below the empirical benchmarks. No single condition matches all three benchmarks, meaning that individual responses are sensitive to the specific political statements included in the persona prompt.

\begin{figure}[t]
    \centering
    \includegraphics[width=1.0\linewidth]{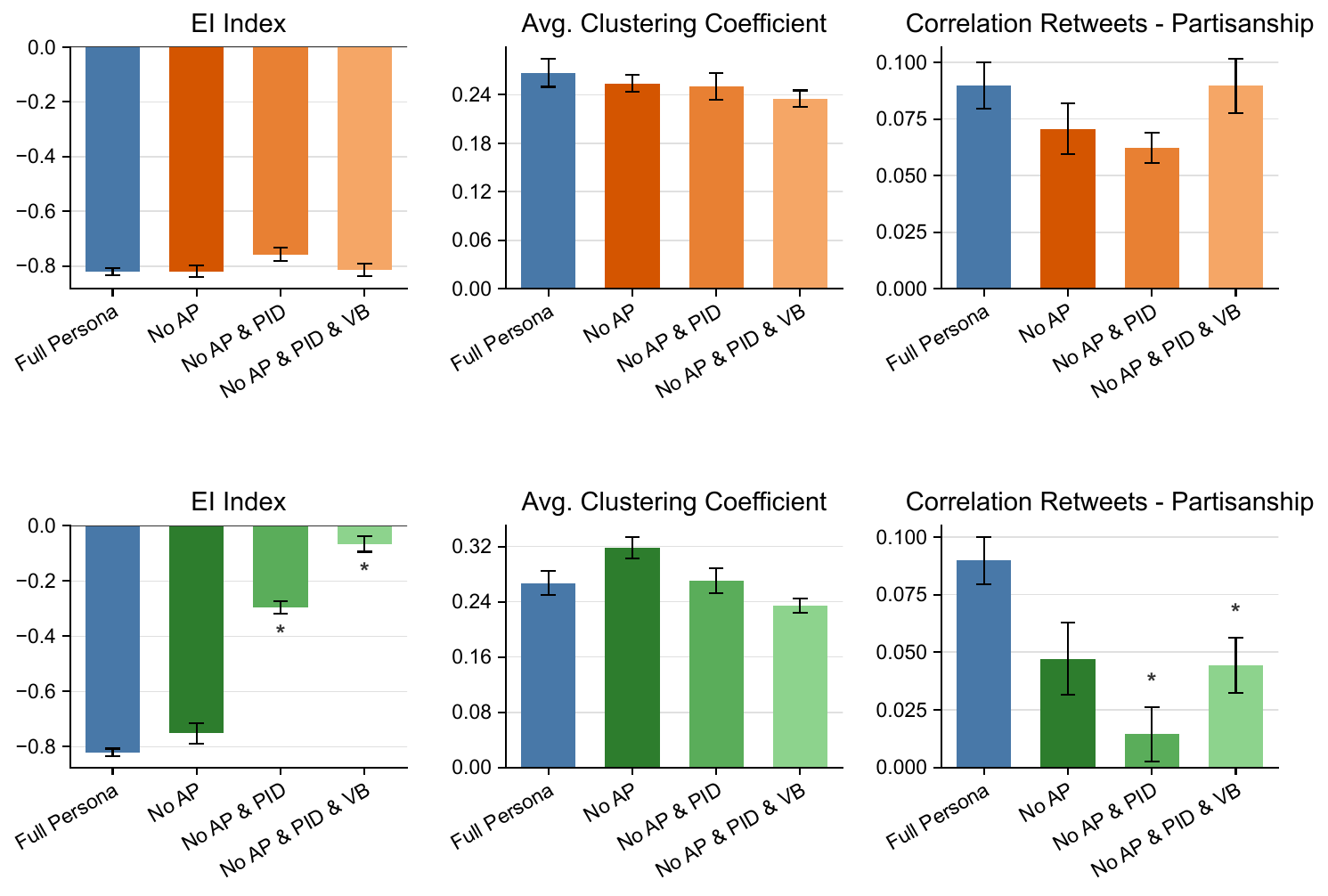}
    \caption{Comparison of network outcomes when modifying public-facing biographies (top) and internal persona prompts (bottom). Public bios do not change the network. Internal personas drive collective polarization, moving the network from near-complete segregation toward integration, with the largest shift after removing AP and PID (Mann-Whitney, $p < 0.05$). They also reduce the correlation between absolute partisanship and received retweets.}
    \label{fig:metrics_comparison}
\end{figure}

\subsection{RQ2: Internal persona specifications drive collective polarization}
If the individual-level distortions identified in RQ1 transform into collective artifacts, removing the corresponding persona statements should alter them. By collective artifact we refer to the outcomes that arise from the persona specification, instead of the interaction dynamics. In the mechanism ablation, we test whether these effects emerge from the public identity profile agents share with each other, or in how they are instructed to behave. \Cref{fig:metrics_comparison} presents the results across three network metrics under each ablation condition. The top panels show ablations applied only to public biographies, while the bottom panels show ablations applied to the persona prompts.
 
\textbf{Biography ablations produce small effects.}
Removing polarizing statements from public biographies slightly shifts platform dynamics. The E-I index remains consistently negative across all biography ablation conditions, indicating that agents continue to primarily connect with co-partisans. The average clustering coefficient shows no statistically significant change but a trend toward reduction, and the retweet-partisanship correlation shows high variance across runs. 
 
\textbf{Persona ablations produce substantial effects.}
In contrast, ablating the internal persona prompts leads to significant changes in collective network outcomes. As polarizing statements are progressively removed, the E-I index shifts significantly upward, indicating a substantial increase in cross-party connections. The removal of AP markers and PID together produces the largest statistically significant shift (Mann-Whitney test, $p < 0.05$), with the network transitioning from complete segregation toward almost full integration. Appendix Fig. 1 illustrates the networks.

The average clustering coefficient shows no statistically significant change in either experiment or model (see other model results in Appendix Fig. 3). The internal ablation leads to a shift toward a lower correlation between retweets and partisanship. Under the full persona, extreme partisan content of either party receives more retweets. After ablation, this relationship weakens.

\begin{figure}[t]
    \centering
    \includegraphics[width=1.0\linewidth]{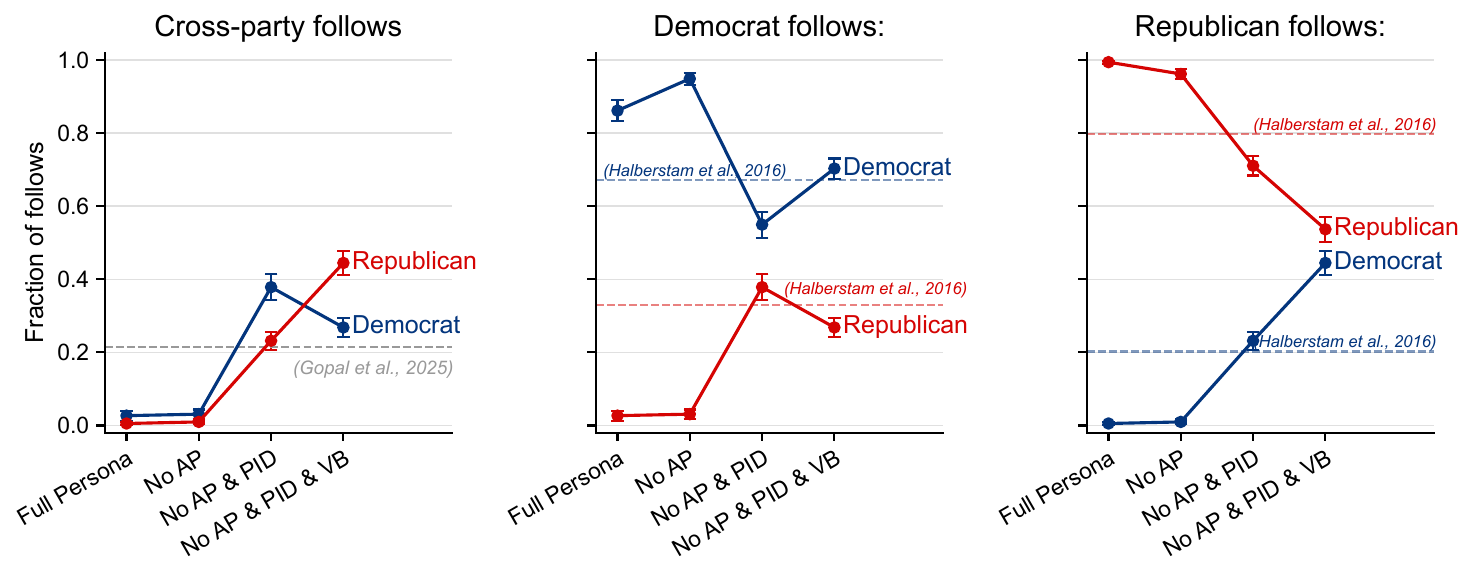}
    \caption{Post-simulation cross-party follow rates compared with Twitter networks of US legislators (left, 21.5\% \cite{gopal2025national}) and of users who follow politicians (middle, right; 33\% of liberals and 20\% of conservatives \cite{halberstam2016homophily}). Ablated personas approach empirical cross-party follow rates. Under the full persona, virtually no cross-party ties form. Removing AP and PID produces the largest increase. No single condition matches all benchmarks.}
    \label{fig:actual_cross_party_follows}
\end{figure}
\subsubsection{No configuration matches all empirical benchmarks.}
We compare cross-party follow rates across ablation conditions with empirical observations from two studies of real political networks on Twitter \cite{halberstam2016homophily,gopal2025national}. In \Cref{fig:actual_cross_party_follows} we show this comparison.
 
Under the full persona, virtually no cross-party ties form (\Cref{fig:actual_cross_party_follows}, left panel). Cumulative ablations increase cross-party ties towards the empirically observed rate of 21.5\% reported by Gopal et al. \cite{gopal2025national}. For Republican agents, removing affective polarization and party identification is sufficient to approach this empirical benchmark, while for Democratic agents, the additional removal of voting behavior is required. We show the effect of each ablation separately in Appendix Fig. 2.
 
The middle and right panels of \Cref{fig:actual_cross_party_follows} compare party-specific follow rates against the findings of Halberstam and Knight \cite{halberstam2016homophily}. The same pattern appears in party-specific rates, with full personas producing agents that follow almost exclusively like-minded agents. As ablations are applied, cross-party following increases toward the empirical benchmarks. For both parties, the removal of AP and PID produces the largest gains. For Republican agents specifically, the setting that best approximates the empirical finding retains voting behavior but removes AP and PID, suggesting that these two components are the primary drivers of the exaggerated homophily under the full persona.

Our results show that different ablations have different outcomes, which sometimes reach empirical benchmarks and sometimes do not. Removing affective polarization lowers the thermometer gap to the ANES survey values, but leaves trait gaps, follow gaps and cross-party ties in the simulation unchanged. Removing party identification as well brings the cross-party following of Republican agents to the empirical benchmark, while Democratic agents also require the removal of voting behavior, at which point the thermometer gap has already fallen below the survey values. No single ablation satisfies all outcomes, and even when a condition matches an empirical result, it is on the surface: the asymmetry between Republican and Democratic agents resembles observational data \cite{halberstam2016homophily}, but contradicts experimental evidence \cite{mosleh2021shared}. These contradictions become visible only by performing ablations in both questionnaire and simulation, instead of trusting the first specification that approaches an empirical finding. Ablations should be used to measure the sensitivity of the conclusions to the persona specification, and not to find the correct persona.

\begin{figure}[t]
    \centering
    \includegraphics[width=1.0\linewidth]{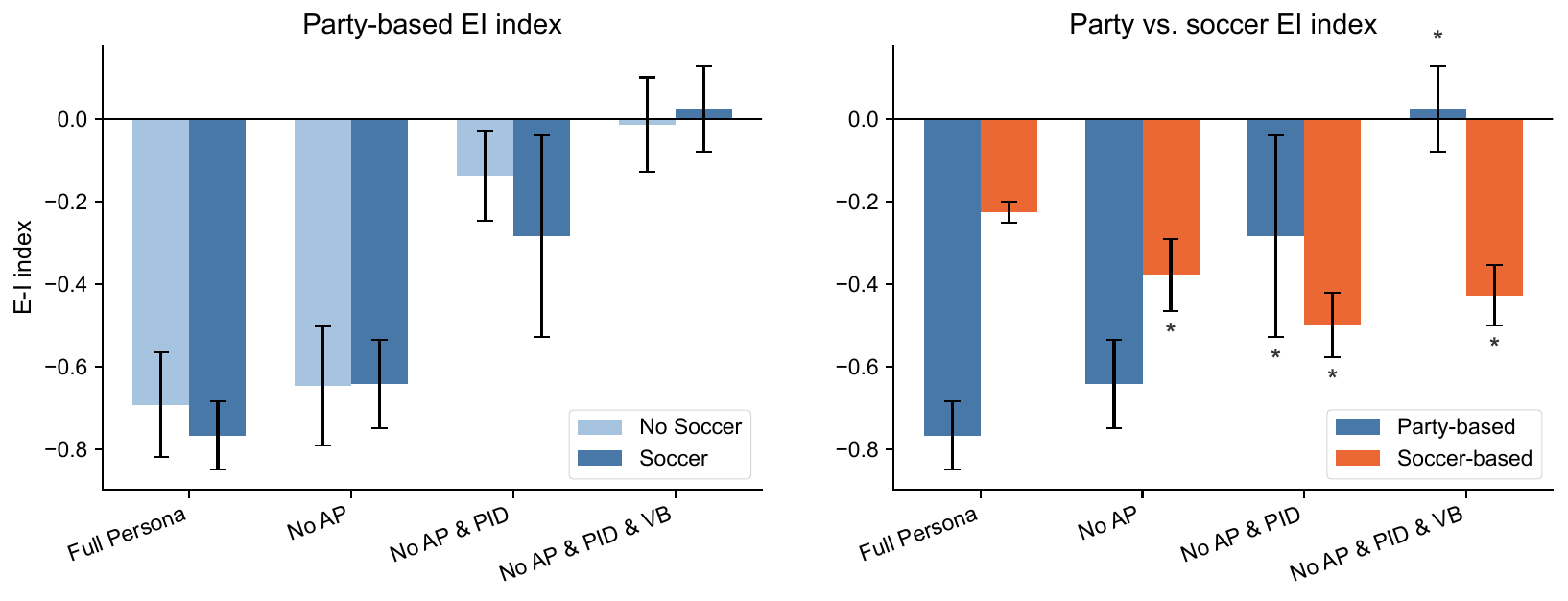}
    \caption{We add "You love soccer" or "You hate soccer" to each persona, evenly split. \textit{Left}: The soccer statements do not change political polarization in any ablation condition. \textit{Right}: As political statements are removed, soccer-based polarization increases.}
    \label{fig:soccer_saliency}
\end{figure}

\subsection{RQ3: Adding non-political statements does not reduce political polarization}
Our previous results show that explicit political identity receives disproportionate behavioral weight. We test whether adding non-political statements reduces the influence of political ones by prompting half of the agents with "You love soccer", while the other half receive "You hate soccer".  In \Cref{fig:soccer_saliency}, we compare the simulation outcomes before and after the injection (left). Across all ablation conditions, political polarization remains unchanged after introducing the soccer statement.

We also compare political polarization with soccer polarization between agents who love and hate soccer (right). As political statements are ablated, political polarization decreases, while soccer-based polarization increases, though not monotonically. This is consistent with the salience explanation. As political statements are removed, the remaining statements gain more weight. However, soccer-based polarization plateaus quickly, and further ablations do not affect it. It also never reaches the levels we observe in political polarization.

\section{A Checklist for Persona Reporting in Generative Social Simulation}
\label{sec:checklist}

GSS's credibility heavily depends on the design of the persona. We suggest the following framework to implement in practice.

\begin{enumerate}
    \item \textbf{Publish the persona prompt verbatim, including the mapping from source data.} Report not only the template or a vague algorithm, but the thresholds and rules that generated each statement.
    \item \textbf{Separate internal personas from information other agents see, and report both.} These are distinct artifacts with distinct functions. One specifies an agent's policy, the other specifies what other agents observe about the agent. They are not interchangeable: in our case, ablating the public biography does not change collective outcomes, while ablating the persona does significantly.
    \item \textbf{State which attributes are behavioral instructions.} Personas from surveys assign attributes equal weight. Authors should report whether that is intended, and whether any attribute is meant to be latent.
    \item \textbf{Perform mechanism ablations on personas.} Report both the marginal and cumulative effects of each component, with statistical tests across seeds. An ablation should be treated as a control condition.
    \item \textbf{Test alternative persona formulations.} Vary the persona's representation, while holding the data and environment settings fixed, and report whether the conclusion remains. Ye et al. show that personas stated as prose versus a bullet list can change cooperation rates by up to 76 percentage points \cite{ye2026stop}. 
    \item \textbf{Benchmark against several empirical findings.} When available, compare against multiple empirical observations and report which observation the simulation matches and which it does not. Matching only one outcome is not enough for behavioral validity.
    \item \textbf{Report model sensitivity across multiple models.} Report both the sign and the magnitude of the effect, to test whether the effect is robust across multiple different models and not just an artifact of a specific family.
    \item \textbf{Vary the component you want to draw conclusions about.} A claim about platform architecture needs the architecture varied and the personas held fixed. In our experiments the architecture was fixed and the personas alone drove the conclusions.
\end{enumerate}

\section{Conclusion}
We examined how survey-driven personas shape polarization in generative social simulation at the individual and collective level. Explicit partisan markers amplify specific dimensions of stated political behavior. In controlled simulation ablations, we found that removing polarizing statements from the persona reduces collective polarization, while removing them only from public biographies has little effect. Adding non-political statements does not reduce the influence of partisan identity on collective behavior.

Our results suggest that collective outcomes in GSS can be artifacts of how survey attributes are translated into personas and gain disproportionate influence in the agent's behavior. We point to salience as one possible explanation. Survey responses do not report how much respondents care about issues. Persona statements, on the other hand, treat every response as a behavioral instruction. Our analysis shows the limitations of surveys as persona construction sources. Personas built from interviews, where people only voice what matters to them, may reduce the problem \cite{park2024generative}. We provide direct empirical support for recent concerns about the epistemic reliability of LLM-based social simulation \cite{li2026position,touzel2026,ye2026stop}, and suggest concrete steps to detect this sensitivity in future studies. 

We do not suggest that personas are the only source of distortion in GSS, or that ablating them is sufficient for valid collective dynamics. Distortions may also arise from the way interactions and communications are performed in the simulation, or even from the LLM itself. We did not explore the interaction of these mechanisms, so they may introduce further limitations of GSS. For example, Ye et al. find that recommendation feed size can strengthen echo-chamber outcomes while the memory window does not \cite{ye2026stop}.
 
Generative social simulation could be a promising tool for complementing human-subjects research \cite{anthis2025position}. Applying such a tool, however, requires understanding how decisions to construct agents affect both individual behavior and emerging collective dynamics. Future studies should follow rigorous frameworks, in which persona construction, interaction and communication protocols are subjected to sensitivity analysis and benchmarking.

\section*{Acknowledgments}
This work is supported by ERC grant (RE-LINK, 101116987, \url{https://doi.org/10.3030/101116987}). Views and opinions expressed are however those of the author(s) only and do not necessarily reflect those of the European Union or the European Research Council Executive Agency. Neither the European Union nor the granting authority can be held responsible for them.

%
%
\bibliographystyle{splncs04}
\bibliography{bibliography}

\end{document}